\documentclass[12pt]{iopart}
\pdfoutput=1

\usepackage{iopams} 
\usepackage{amssymb, wasysym}
\usepackage{graphicx}
\usepackage{color}
\usepackage{tabularx}
\newcolumntype{C}[1]{>{\centering\arraybackslash}p{#1}}

\usepackage{cite}

\usepackage{longtable}		
\usepackage{marvosym}
\usepackage{multirow}
\usepackage{multicol}
\usepackage{booktabs}

\usepackage{bbm}
\usepackage{rotating}
\newcolumntype{C}[1]{>{\centering\arraybackslash}p{#1}}

\usepackage{longtable}		
\usepackage{marvosym}
\usepackage{array}
\usepackage{verbatim}

\definecolor{orange}{rgb}{1,0.65,0}

\usepackage{pifont}
\usepackage{ifsym}
\usepackage{textcomp}
\usepackage{units} 
\usepackage{caption}[2008/08/24]

\usepackage[english]{babel}

\usepackage[hyphens]{url} 

\usepackage{lmodern} 
\usepackage[T1]{fontenc}

\definecolor{orange}{rgb}{1,0.65,0}

\newcommand{\text}[1]{#1}

\newcommand{\eqref}[1]{(\ref{#1})}
\newcommand{\V}{Vy\-cor\ }

\newcommand{\dC}{\,$^{\circ}$C}

\newcommand{\subs}[2]{#1_{\mbox{{\scriptsize #2}}}}

\newcommand{\bild}[1]{Fig.~\ref{#1}}

\newcommand{\rel}[1]{Eq.~\eqref{#1}}

\begin{document}

\title[Imbibition in Mesoporous Silica]{Imbibition in Mesoporous Silica: Rheological Concepts and Experiments on Water and a Liquid Crystal}

\author{Simon Gruener$^{1,2}$ and Patrick Huber$^1$}
\address{$^1$ Experimental Physics, Saarland University, D-66041 Saarbr\"{u}cken, Germany}
\address{$^2$ Department of Mechanical and Aerospace Engineering, Princeton University, Princeton, NJ 08544, USA}
\eads{\mailto{sgruner@princeton.edu}, \mailto{p.huber@physik.uni-saarland.de}}

\begin{abstract}
Along with some fundamental concepts regarding imbibition of liquids in porous hosts we present an experimental, gravimetric study on the capillarity-driven invasion dynamics of water and of the rod-like liquid crystal octyloxycyanobiphenyl (8OCB) in networks of pores a few nanometers across in monolithic silica glass (Vycor). We observe, in agreement with theoretical predictions, square-root of time invasion dynamics and a sticky velocity boundary condition for both liquids investigated. Temperature-dependent spontaneous imbibition experiments on 8OCB reveal the existence of a paranematic phase due to the molecular alignment induced by the pore walls even at temperatures well beyond the clearing point. The ever present velocity gradient in the pores is likely to further enhance this ordering phenomenon and prevent any layering in molecular stacks, eventually resulting in a suppression of the smectic in favor of the nematic phase.
\end{abstract}

\maketitle

\section{Introduction}
Liquid flow propelled by capillary forces is one of the most important transport mechanisms in porous environments. It is governed by a fascinating interplay of interfacial, viscous drag as well as gravitational forces which liquids encounter upon invasion into geometries with often complex topologies, such as capillary networks of trees \cite{Bear1988, Koch04, Wheeler08} or interconnected fractures in soils and ice. There has been significant progress with regard to a quantitative understanding of this phenomenon, most prominently with regard to the transport through rocks and soils,  starting with the seminal work of the French engineer Henry Darcy \cite{Sahimi1993, Mecke2005}. Almost all experimental and theoretical studies, published to date, deal however with porous structures where the characteristic length scales, meaning the typical pore diameters, are on the macroscale, that is much larger than the typical size of the basic building blocks (molecules) of the liquids \cite{Alava04}.

After an introduction of the theoretical and experimental fundamentals, exemplified for the case of water imbibition in mesopores, we present a temperature-dependent imbibition study on the liquid crystal octyloxycyanobiphenyl (8OCB). The goal is to scrutinize which principles regarding its rheology, known from the bulk state, can be safely transferred to the transport across nano- or mesoscopic geometries. Finally, insights into the phase transition behavior of the spatially nano-confined, rod-like liquid crystal will be gained.

Given the emerging interest in micro- and nanofluidic applications \cite{Sparreboom09} such a study on the non-equilibrium behavior of liquids in extreme spatial confinement is not only of fundamental interest, but also of rather practical importance \cite{Eijkel05, Beugelaar10, Arora10}. Moreover, the increasing use of mesoporous matrices as hard templates \cite{Thomas08} for the preparation of well-defined nanoscopic, soft-matter structures, such as nanotubes and nanorods \cite{Luo04}, necessitates a profound understanding of this phenomenology.

\section{Fundamentals of liquid imbibition in mesoporous solids}    \label{sectionFundamentals}

\subsection{Liquid flow in isotropic pore networks: Darcy's law}  \label{sectionDarcysLaw}
In order to introduce the fundamentals of liquid flow in mesoporous media, we start with a much simpler phenomenology: The flow of a liquid through a single straight pipe. According to the law of Hagen-Poiseuille the volume flow rate $\dot{V}_{\rm HP}$ for a given pressure difference $\Delta p$\label{pressuredifference} applied along the cylindrical duct with radius $r$\label{radius1} and length $\ell$ is determined by
\begin{equation}
\dot{V}_{\rm HP} = \frac{\pi \,r^4 }{8\,\eta \,\ell} \,\Delta p \; .
\label{eq:HagenPoiseuille}
\end{equation}
Here $\eta$\label{viscosity} denotes the dynamic viscosity of the liquid. In the following we will present a simple concept that allows one to extend the applicability of \rel{eq:HagenPoiseuille} to the flow through a complex pore network \cite{Bear1988}.

Porous Vycor glass (code 7930) provided by Corning Incorporated was used over the course of this study. It is produced through metastable phase separation in an alkali-borosilicate system followed by an extraction of the alkali-rich phase. The final glass consists of a sponge-like network of tortuous and interconnected pores with mean pore radii on the nanometer scale embedded in a matrix mostly consisting of SiO$_2$ \cite{Elmer92}. 

Such an isotropic pore network can be characterized by a set of three quantities. The pore radius $\subs{r}{0}$\label{meanporeradius} and the volume porosity $\subs{\phi}{0}$\label{volumeporosity} are probably the most intuitive ones among them. Two different batches of Vycor glass were applied in this study that differed in $\subs{r}{0}$ whereas they coincided in $\subs{\phi}{0}= 0.32 \pm 0.01$. We will refer to them as V5 [$\subs{r}{0} = (3.4\pm 0.1)$\,nm] and V10 [$\subs{r}{0} = (4.9 \pm 0.1)$\,nm] from now on \cite{gruener10a}. 

These matrix properties were accurately ascertained employing nitrogen sorption isotherms conducted at 77\,K and through a subsequent ana\-ly\-sis within a mean field model for capillary condensation in nano- and mesopores \cite{Saam75}. We obtained pore size distributions $P(r)$ with standard deviations of approximately $0.1\,r_0$ (with the peak or most probable value of the distribution $r_0$). In order to assess the impact of this uncertainty in $r$ we turn to the quantity measured in our experiments that is the mass increase of the porous matrix due to the uptake of a liquid -- see section~\ref{experimental} for details. Here each pore gives a signal that is proportional to its volume $\propto r^2$ and consequently the macroscopic signal, which is averaged over the complete sample, should be identically equal to the result for an imaginary porous host with the single pore radius
\begin{equation}
r' \equiv \sqrt{\frac{\int \, r^2\,P(r)\,{\rm d}r}{\int \,P(r)\,{\rm d}r}} \; .
\label{eq:psd}
\end{equation}
Since we could not find any difference between $r_0$ and $r'$ exceeding their error margins we conclude that the distributions can be considered to be narrow and negligible in these specific measurements. Hence we proceed taking into consideration only the most probable pore radii $r_0$ stated above.

To account for the isotropy of the network the so-called tortuosity $\tau$\label{tortuosity} must be introduced along with the transformation $\dot{V}  \rightarrow  \tau^{-1}\, \dot{V}$. Pores totally aligned in flow direction yield $\tau=1$, whereas isotropically distributed pores result in $\tau=3$ since for randomly oriented channels only every third pore is subjected to the pressure gradient and hence contributes to the flow. To date many different methods have been applied in order to extract the tortuosity of porous \V glass \cite{Lin92, Bommer08, Crossley91}. They consistently yield $\tau \approx 3.6$ which, interestingly, deviates from $\tau=3$. This is because the pores are not straight but rather meandering. A value $\tau >3$ corrects the pore length for the larger flow path.

Finally one is able to derive an expression that describes the flow of a liquid through a porous network. For a given porous matrix with cross-sectional area $A$ and thickness $d$\label{samplethickness} (along which the pressure drop $\Delta p$ is applied) the volume flow rate $\dot{V}_{\rm D}$ is determined by Darcy's law \cite{Debye59}
\begin{equation}
 \dot{V}_{\rm D} = A\, \frac{\subs{\phi}{0}\, r_0^2}{8\,\tau\,\eta\,d} \,\Delta p \; .
\label{eq:Darcy}
\end{equation}

\subsection{Velocity boundary conditions}  \label{subsectionNSBC}
The law of Hagen-Poiseuille implies the no-slip boundary condition. This means that the velocity of the fluid layers directly adjacent to the restricting walls equal the velocity of the walls themselves. Nowadays it is indisputable that this assumption does not hold unreservedly. Already 60 years ago Peter Debye and Robert Cleland introduced both slipping and sticking fluid layers at the pore walls in order to interpret a seminal experiment on liquid flow across porous \V \cite{Debye59}.

Up to date many factors have been found that seem to influence the boundary conditions \cite{Granick03, Lauga05, Neto05}. The least controversially discussed amongst them is the fluid-wall interaction \cite{Barrat99, Pit00, Cieplak01, Tretheway02, Cho04, Schmatko05, Fetzer07, Baeumchen10, Voronov08, Maali08, Servantie08, Sendner09}. Shear rates beyond a critical value are supposed to induce slip, too \cite{Zhu01, Craig01, Priezjev04, Priezjev07}. In contrast, the influence of surface roughness is rather debatable \cite{Vinogradova06, Zhu02, Pit00, Bonaccurso03}. Furthermore, dissolved gases \cite{Granick03, Dammer06}, the shape of the fluid molecules \cite{Schmatko05} or the add-on of surfactants \cite{Cheikh03} might influence the boundary conditions as well. 

\begin{figure}[!ht]
\centering
\includegraphics*[width=.5\linewidth]{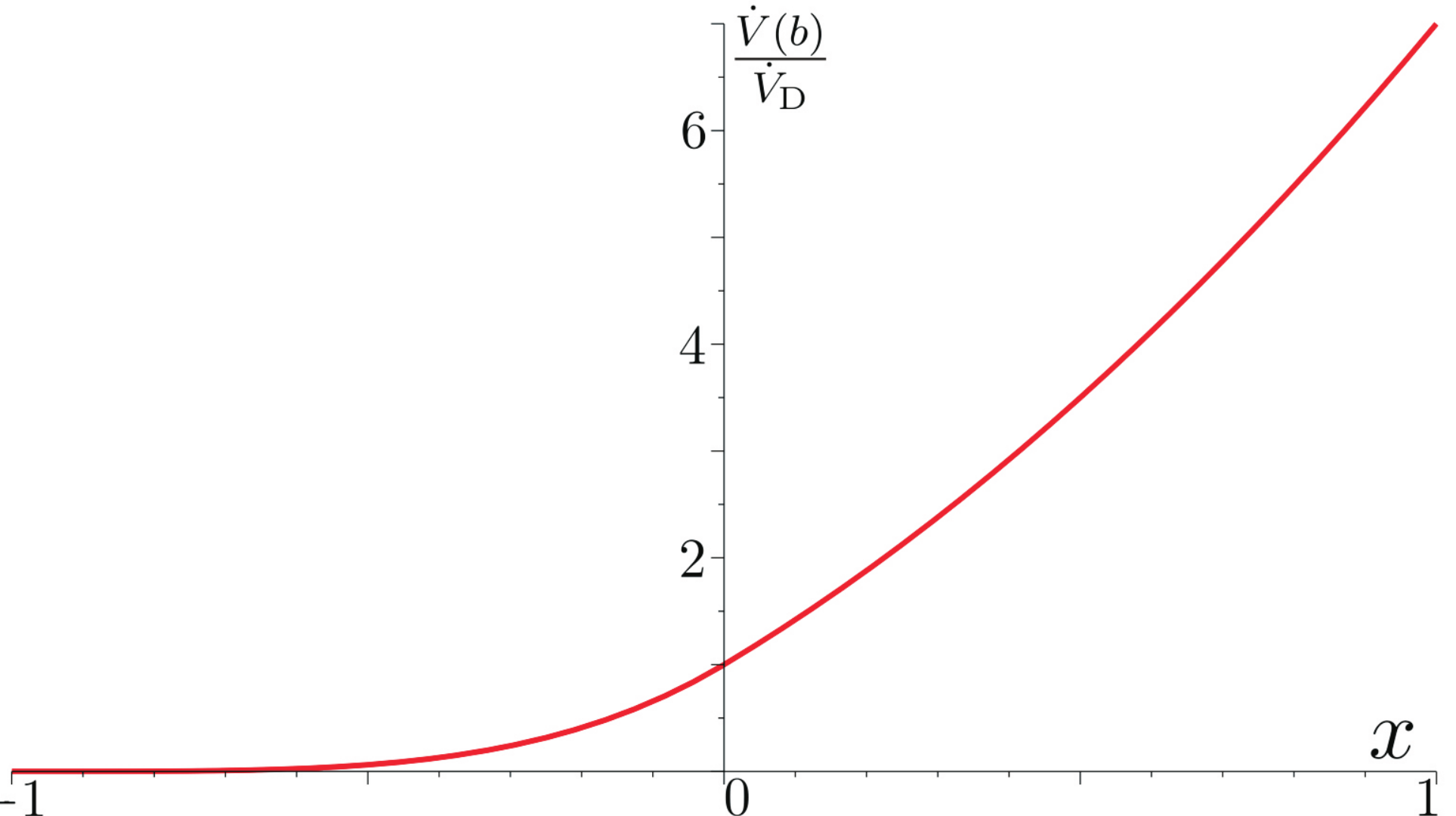}
\caption{Volume flow rate $\dot V (b)$ normalized with Darcy's flow rate $\dot V_{\rm D}$ \rel{eq:Darcy} as a function of the slip length $b$ in units of the pore radius $r_0$: $x\equiv\frac{b}{r_0}$. Note that the flow rate vanishes for $b \to -r_0  \,\Leftrightarrow \, x \to -1$.}
\label{flowrate}
\end{figure} 
The degree of slip can be quantified by the slip length $b$\label{sliplength} with $\subs{r}{h} \equiv \subs{r}{0}+b$ the hydrodynamic pore radius\label{hydrodynamicradius}, which measures the distance from the pore center to the radius where the streaming velocity reaches zero. Thus, a sticking layer boundary condition corresponds to $b<0$. The volume flow rate $ \dot{V} (b)$ can be calculated from the parabolic velocity profile 
\begin{equation}
v_{\rm z}(r)= \frac{\Delta p }{4\,\eta\,\ell} \; \left[ (r_0+b)^2 -r^2  \right]
\label{eq:velprof}
\end{equation}
through an integration over the channel's cross-sectional area $\pi\,r_0^2$ if $b>0$. For $b<0$ however the immobile layers have to be excluded. Then $\dot V (b)$ in units of $\dot V_{\rm D}$ is given by
\begin{equation}
\frac{\dot V (b)}{\dot V_{\rm D}} = \left\{
\begin{array}{lcl}
	(x+1)^4 & {\rm if} & b<0 \\
	1+4\,x+2\,x^2 & {\rm if} & b>0	
\end{array} \right.
\label{eq:normflowrate}
\end{equation}
in terms of normalized slip lengths $x\equiv \frac{b}{r_0}$. In \bild{flowrate} this quantity is plottet for $-r_0 < b < r_0$. Because of the obvious bijectivity of $\dot V (b)$ any measured flow rate $\dot V$ can unambiguously be related to a unique slip length $b$. Part of the evaluation of the present investigation is based upon this principle.

\subsection{Dynamics of spontaneous imbibition}
From the physisist's point of view spontaneous imbibition is an impressive example for interfacial physics. The driving force behind the capillary rise process is the Laplace pressure $\subs{p}{L}$\label{Laplacepressure} acting on the curved meniscus of a liquid in a pore or porous structure. It is specified by  
\begin{equation}
\subs{p}{L} = \frac{2\,\sigma\,\cos\subs{\theta}{0}}{\subs{r}{0}}
\label{eq:Laplace2}
\end{equation}
with $\sigma$\label{surfacetension} the surface tension of the liquid. This implies that spontaneous imbibition can only occur for a wetting or partially wetting liquid.

As the liquid rises beyond its bulk reservoir to a certain level $h$ the hydrostatic pressure $\subs{p}{h}= \rho\,g\,h$ acting on the liquid column increases. The final state is derived from a balance between $\subs{p}{L}$ and $\subs{p}{h}$ \cite{Caupin08} resulting in the maximum rise level given by Jurin's law. However, because of the tiny pore diameters considered in our study typical Laplace pressures are on the order of a few 100\,bar and hence much higher than typical hydrostatic pressures. The gravitational force may therefore be neglected.

\begin{figure}[!t]
\centering
\includegraphics*[width=.5\linewidth]{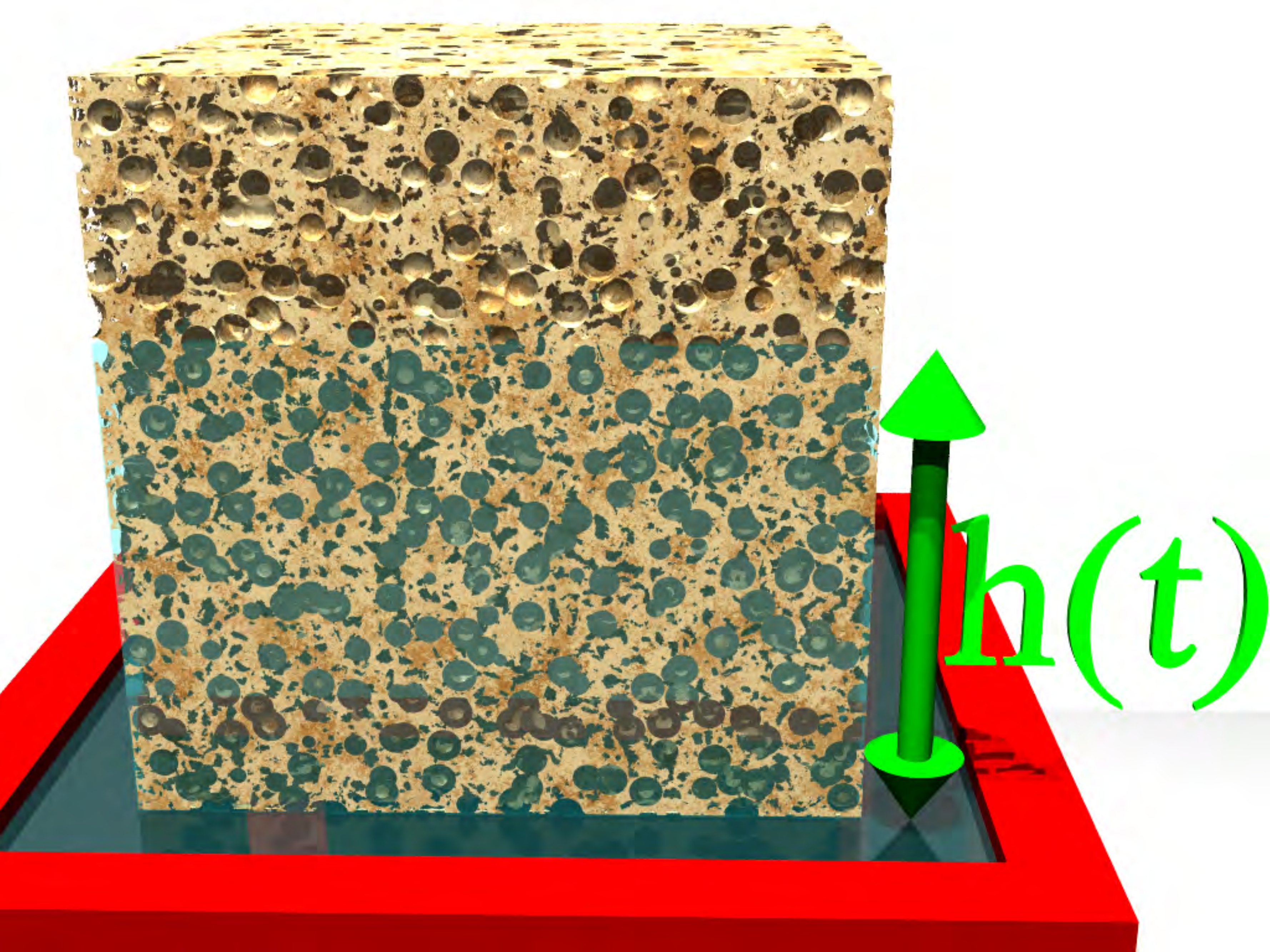}
\caption[\V sample during a capillary rise experiment]{Raytracing illustration of a \V sample during a capillary rise experiment filled up to the level $h(t)$.}
\label{Vycor_filled}
\end{figure}

In order to gain information on the dynamics of the imbibition process in a porous network we can now resort to Darcy's law. Considering Fig.~\ref{Vycor_filled}
one easily concludes that the sample height $d$ appearing in \rel{eq:Darcy} has to be replaced by the time-dependent rise level $h(t)$ since only the filled parts of the sample contribute to the flow dynamics. Moreover, at any given time $t$ both the imbibed fluid volume $V(t)$ and its mass $m(t)$ are closely related to $h(t)$ (utilizing the fluid's density $\rho$) via
\begin{equation}
V(t)= \frac{m(t)}{\rho} =\phi_0 \,A\, h(t)\; .
\label{eq:LW1}
\end{equation}
Eventually this leads to the simple differential equation
\begin{equation}
\dot{h}(t)\, h(t) = \frac{r_0^2}{8\,\tau\,\eta} \,\Delta p
\label{eq:LW2}
\end{equation}
solved through
\begin{equation}
h(t) = \sqrt{\frac{r_0^2}{4\,\tau\,\eta } \,\Delta p} \; \sqrt{t}\; 
\label{eq:LW}
\end{equation}
the Lucas-Washburn (LW) law \cite{Lucas18, Washburn21}. Substituting the Laplace pressure for $\Delta p$ and utilizing \rel{eq:LW1} finally yields
\begin{equation}
m(t) = \underbrace{\rho\,A \,\phi_0\,\sqrt{\frac{r_0\,\sigma}{2\,\tau\,\eta}}}_{C} \;\sqrt{t} \; ,
\label{eq:imb_mt}
\end{equation}
the mass increase $m(t)$ of the sample due to the liquid uptake as a function of the time\footnote{Since all liquids considered in our study totally wet the high-energy silica surface of the pore walls (as confirmed by contact angle measurements in the group of Karin Jacobs at Saarland University, Saarbruecken, Germany) the contact angle is already eliminated because of $\cos(0^\circ)=1$.}. It is noteworthy that molecular dynamics simulations \cite{Gelb02, Binder07} corroborate on a microscopic scale the validity of this model for imbibition of molecular fluids as derived from macroscopic, rather phenomenological considerations even down to meso- and nanopores.

The imbibition coefficient $C$ is a simple measure of the invasion dynamics. Please note that the above equation explicitly obeys the no-slip boundary condition. Hence, any deviation of a measured imbibition coefficient from the one predicted according to \rel{eq:imb_mt} may directly be interpreted in terms of a non-zero slip length $b$ utilizing \rel{eq:normflowrate}.

\section{Experimental Part} \label{experimental}
\begin{figure}[!t]
\centering
\includegraphics*[width=.5\linewidth]{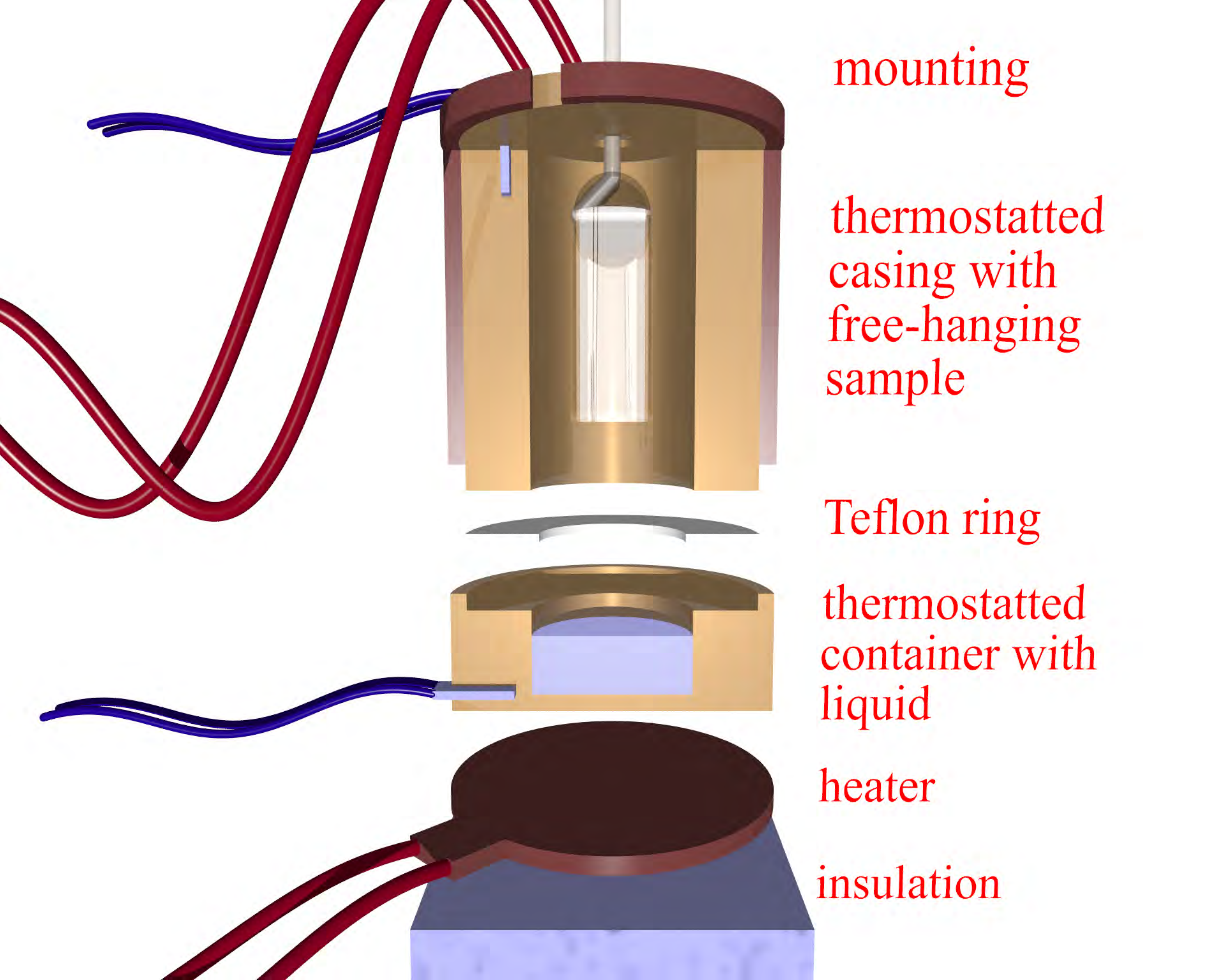}
\caption[Exploded view of the imbibition setup (LIS)]{Exploded view of the imbibition setup (raytracing illustration). The top of the \V sample is glued to a wire and hangs freely into the cell. The cell itself consists of a casing and a container. Both are build out of copper and can separately be thermostatted. The cell itself can be moved in vertical direction.}
\label{LIS_cell}
\end{figure}

Imbibition dynamics are studied via a recording of the samples' mass increase $m(t)$. The setup is depicted in Fig.~\ref{LIS_cell}. For a time-dependent measurement of the force acting on and, hence, of the mass increase of the porous \V block the sample is installed on standard laboratory scales applying a special mounting. In order to perform measurements beyond room temperature a cell was constructed that allows for a simultaneous thermostatting of the sample itself and the liquid reservoir beneath. 

\begin{figure}[!t]
\centering
\includegraphics*[width=.5\linewidth]{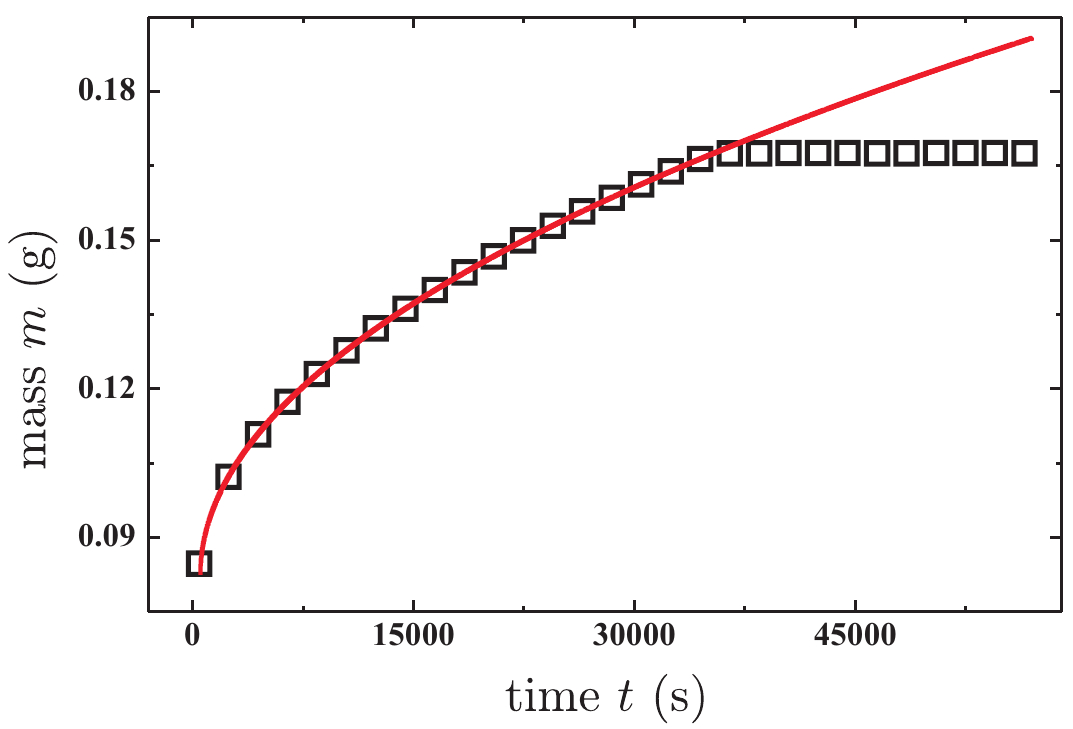}
\caption{Measurement of the mass increase of a porous \V block upon the liquid invasion during an imbibition experiment. The solid line is a $\sqrt{t}$-fit according to Eq.~\eqref{eq:imb_mt}. The data density is reduced by a factor of 2000.}
\label{imb_bsp}
\end{figure} 

Upon start of the experiment the cell is moved upward until the sample touches the liquid surface. The liquid immediately starts invading the porous host what becomes noticeable by the gradual increase of the sample's mass (see Fig.~\ref{imb_bsp}). According to the previously introduced model of imbibition and Eq.~\eqref{eq:imb_mt} this behavior can be very well described by a $\sqrt t$-fit (see solid line in Fig.~\ref{imb_bsp}), which yields the imbibition coefficient $C$. Eventually a level of saturation is entered, which signals that the sample is completely filled. 

\subsection{Data interpretation using the example of mesopore-confined water}  \label{confinedwater}
Experiments on the capillary rise of water in V5 and V10 were conducted at three different temperatures $T$ (25\dC, 40\dC, and 60\dC). The measured flow rates showed significant deviations from the expected value according to \rel{eq:imb_mt}. Utilizing \rel{eq:normflowrate} this discrepancy could be transferred into a non-zero slip length. Consistently for both sample batches a negative slip length of about 5\,\AA\ was found. Considering the diameter of a water molecule of approximately 2.5\,\AA\ this result can be interpreted as follows: two layers of water directly adjacent to the pore walls are immobile meaning that they are pinned and do not take part in the flow. The residual inner compartment of the liquid however obeys classical hydrodynamics in accordance with former findings on the conserved fluidity \cite{Israelachvili86, Horn89, Raviv01} and capillarity \cite{Fisher81, Fradin00} of confined water.                                                    

The idea of the compartmentation of mesopore-confined water is corroborated by recent molecular dynamics studies on the glassy structure of water boundary layers in \V and the expected existence of sticky boundary layers in Hagen-Poiseuille nanochannel flows for strong fluid-wall interactions \cite{Heinbuch89, Ricci00, Gallo00, Vichit00, Gallo01, Castrillon09}. By means of X-ray diffraction distortions of the hydrogen-bonded network of water near silica surfaces were found \cite{Fouzri02}, which might be responsible for the markedly altered liquid properties. Tip-surface measurements document a sudden increase in the viscosity by orders of magnitude in 0.5\,nm proximity to hydrophilic glass surfaces \cite{Li07, Khan10, Xu10}. It also extends former experimental results with respect to the validity of the no-slip boundary condition for water/silica interfaces, in which this condition was proven down to at least 10\,nm from the surface \cite{Lasne08} whereas slip-flow for water is only expected at hydrophobic surfaces \cite{Vinogradova99, Lasne08}. For more detailed information we would like to refer the interested reader to two publications reporting on capillary rise dynamics of water in mesoporous silica \cite{Gruener09, Huber07}.

It is interesting to note that such a compartmentisation of the pore fluid with respect to its dynamics has also been inferred for the linear n-alkanes confined in mesopores, both from imbibition experiments \cite{Gruener09a} and from neutron-spin echo experiments on the self-diffusion dynamics \cite{Kusmin10b}.

\section{Rheology and phase transition behavior of the liquid crystal 8OCB in mesopore-confinement}
Usually phase transitions are accompanied by unique variations of the fluid properties. Since the capillary rise dynamics, and in particular the imbibition coefficient $C$, sensitively depend on these quantities one can easily detect such characteristic deviations by performing measurements at different temperatures in the $T$-range of interest. It is convenient to normalize the extracted prefactors $C(T)$ with the value $C_{\rm n} \equiv C(T_{\rm n})$ at an arbitrarily chosen temperature $T_{\rm n}$. This procedure eliminates all quantities that do not depend on the temperature. Particularly, the geometry ($A$) as well as the internal topology of the substrate ($r_0$, $\phi_0$, and $\tau$) do not play a role any longer. Only the fluid properties $\sigma(T)$, $\eta(T)$, and $\rho(T)$ enter this quantity which will henceforth be referred to as normalized imbibition speed $v_{\rm n} (T)$
\begin{equation}
v_{\rm n}(T) \equiv \frac{C (T)}{C(T_{\rm n})} = \frac{\rho(T)}{\rho(T_{\rm n})} \cdot \sqrt{\frac{\sigma(T) \cdot \eta(T_{\rm n})}{\sigma(T_{\rm n}) \cdot \eta(T)}} \; .
\label{eq:normimbspeed}
\end{equation}
One should not be confused by the term `speed', since $v_{\rm n}$ is actually dimensionless. Nevertheless, it quantifies the mass uptake rate of the sample at a given temperature and a characteristic variation in any fluid property must be mapped in this quantity. 

Here we present a thorough study on the capillary rise dynamics of the thermotropic liquid crystal octyloxycyanobiphenyl (8OCB) in mesopores. The building blocks of this liquid may be considered to be rigid rods with a length of $\sim 2$\,nm and diameter between 1\,nm and 1.5\,nm. Bulk 8OCB undergoes a transition from the smectic A to the smectic N phase at $T=67\,^{\circ}{\rm C}$ and beyond the clearing point $\subs{T}{c}=80\,^{\circ}{\rm C}$ any orientational ordering of the molecules disappears (isotropic phase). In particular this last transition is accompanied by the occurrence of a characteristic shear viscosity minimum, what might uniquely influence the invasion dynamics. In the following this behavior will be described in detail.

\subsection{Nemato-Hydrodynamics: Shear viscosity minimum and presmectic divergence of flowing nematic liquid crystals}
The momentum transport in nematic liquid crystals shows an anisotropy since it depends on the mutual orientations of the macroscopic molecular alignment (the director $\vec{n}$), the flow velocity ($\vec{v}$) and the velocity gradient ($\nabla\, v$). In 1946 Mieso\-wicz\label{Miesowicz} defined three principal shear viscosity coefficients of nematics \cite{Miesowicz46}, which can be measured in three different Couette flow experiments sketched in \bild{LC_visc}. Typically magnetic fields are applied in order to align the molecules in the nematic sample. Intuition suggests that the lowest resistance to the nematic flow, i.e. the lowest viscosity value, should be $\eta_2$. Among the two remaining viscosities, $\eta_1$ should have the highest value.
\begin{figure}[!t]
\centering
\includegraphics*[width=.5 \linewidth]{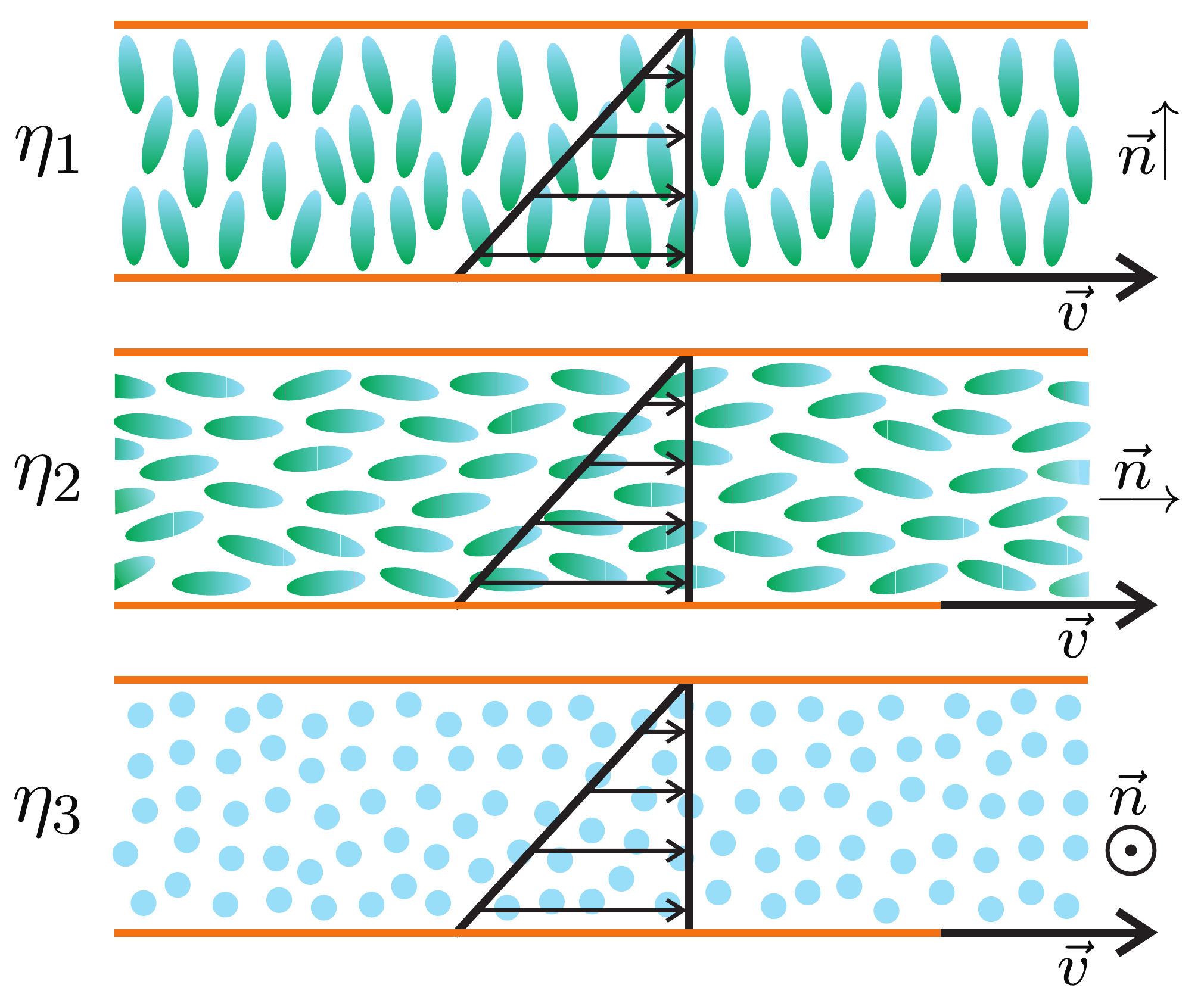}
\caption[Illustration of the three Miesowicz shear viscosity coefficients of nematic liquid crystals]{The experimental Couette flow conditions for measurements of the three Miesowicz shear viscosity coefficients of nematic liquid crystals: $\eta_1$ for $\vec{n}\,\bot\, \vec{v}$ and $\vec{n} \,||\, \nabla \,v$, $\eta_2$ for $\vec{n}\,||\, \vec{v}$ and $\vec{n} \,\bot\, \nabla \,v$, $\eta_3$ for $\vec{n}\,\bot\, \vec{v}$ and $\vec{n}\, \bot\, \nabla \,v$.}
\label{LC_visc}
\end{figure}
\begin{figure}[!t]
\centering
\includegraphics*[width=.5 \linewidth]{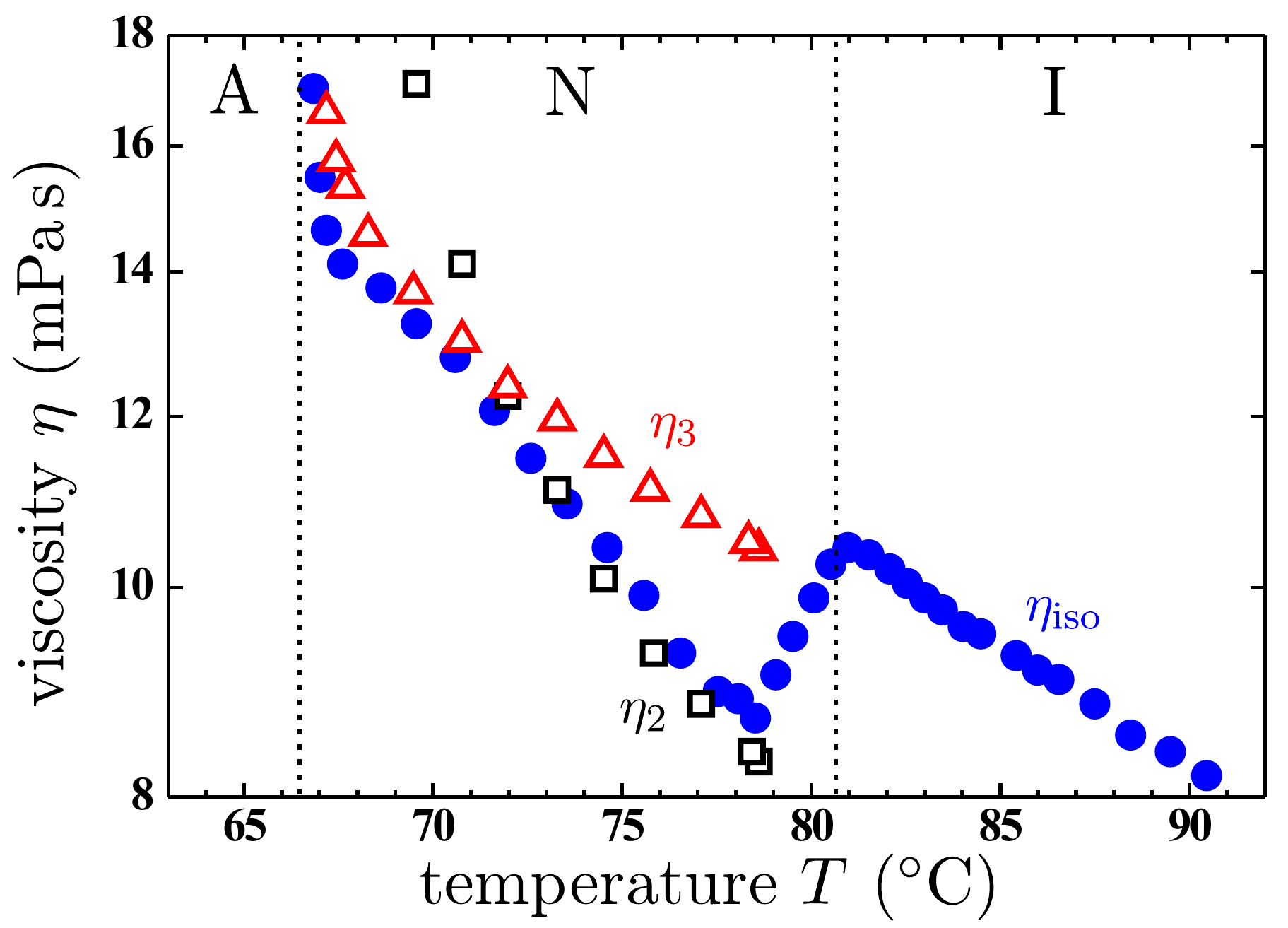}
\caption[Miesowicz and free flow viscosities of 8OCB]{Miesowicz shear viscosities $\eta_2$ ($\square$) and $\eta_3$ (${\color{red} \triangle}$) of the liquid crystal 8OCB compared to its free flow viscosity $\subs{\eta}{iso}$ ({\color{blue} \Circsteel}) according to \cite{Graf92}.}
\label{LC_visc_8OCB}
\end{figure}

Indeed, when the orienting magnetic field, i.e. the director $\vec{n}$, is parallel to the velocity $\vec{v}$ of the nematic flow, the lowest viscosity is recorded (see open symbols in \bild{LC_visc_8OCB}). Nevertheless, this relatively simple picture of the viscosity of nematic liquid crystals is disturbed for the compounds exhibiting the transition to the smectic A phase. Then, with decreasing temperature, the viscosity $\eta_2$ shows a strong increase and goes to infinity at the N-A phase transition. The $\eta_1$ and $\eta_3$ viscosities are almost unaffected. At a temperature that is a few degrees below the temperature at which the transition to the smectic A phase takes place, the viscosities $\eta_2$ and $\eta_3$ interchange their roles and then the lowest nematic viscosity corresponds to the flow in the $\eta_3$ configuration. The presmectic behavior of the $\eta_2$ viscosity is due to the formation of precursors of smectic planes with $\vec{z}\,||\,\vec{v}$ that would be immediately destroyed by the velocity gradient, thus this configuration is rendered unfavorable.

The behavior of the freely flowing compound obeys a general principle that can be formulated as follows: a free fluid adopts such a manner of flow, as corresponds to the minimum of its viscosity at given conditions \cite{Jadzyn01}. Accordingly, the transition from the isotropic to the nematic phase manifests itself in a strong decrease of the shear viscosity $\subs{\eta}{iso}$ that is very close to $\eta_2$. Consistently, beyond the presmectic cross-over of $\eta_2$ and $\eta_3$ the viscosity of the freely flowing liquid crystal $\subs{\eta}{iso}$ follows $\eta_3$ (see filled symbols in \bild{LC_visc_8OCB}). This result is interpreted in terms of rearrangements of the molecular alignment $\vec{n}$ with respect to the velocity field $\vec{v}$, which can easily be assessed by means of examinations of the compound's viscosity.

\subsection{Results and interpretation}
In \bild{LC_massincrease} five measurements of 8OCB invading V5 are shown along with $\sqrt{t}$-fits. Their coincidence is evident and hence the solution of the differential equation \rel{eq:LW2} through a $\sqrt t$-law is justified. This implicitly means that the right-hand side of \rel{eq:LW2} is not a function of the time at all, what has only tacidly been assumed so far. Especially the viscosity consequently is not a function of neither the time nor the time-dependent shear rate, which is related to the rise speed $ \frac{{\rm d}h(t)}{{\rm d}t}\propto \frac{1}{\sqrt{t}}$. Shear thinning or thickening effects of the liquid crystal can therefore be excluded.

The extracted normalized imbibition speeds $v_{\rm n}(T)$ are indicated by the single points in the upper panel of \bild{LC_speeds}. Moreover, utilizing $\subs{\eta}{iso}(T)$-values according to \bild{LC_visc_8OCB} along with the known $T$-dependencies of $\sigma$ and $\rho$ one can also calculate $v_{\rm n}(T)$ (see solid line in the upper panel of \bild{LC_speeds}). 
\begin{figure}[!t]
\centering
\includegraphics*[width=.5 \linewidth]{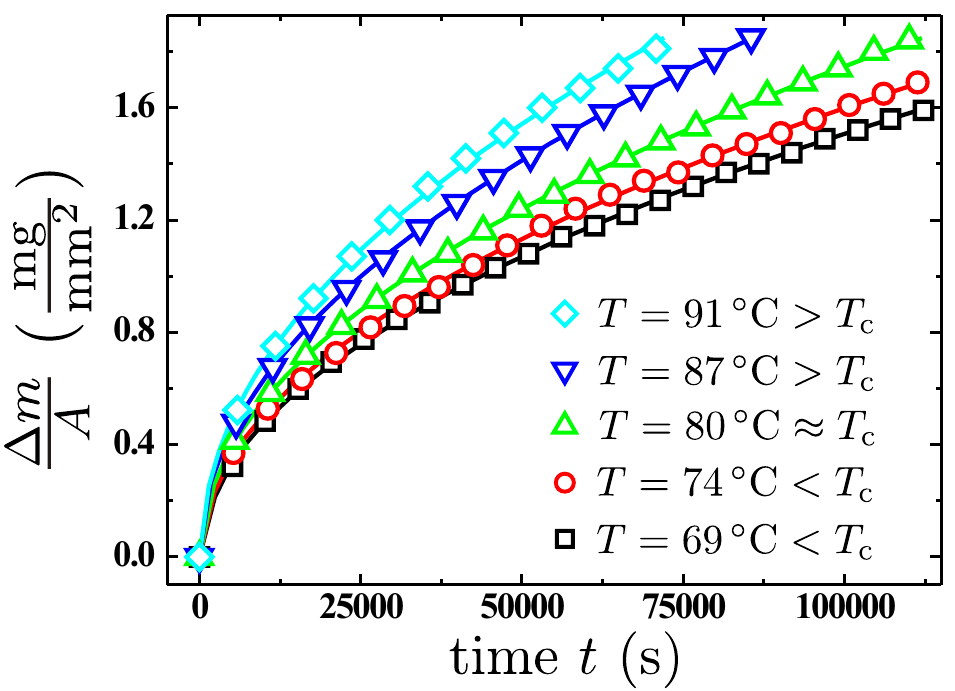}
\caption[Mass uptake of V5 due to liquid crystal imbibition at selected temperatures]{Specific mass uptake of V5 due to the imbibition of the liquid crystal 8OCB as a function of the time for selected temperatures below and above the clearing point $\subs{T}{c} \approx 80$\dC. Solid lines correspond to $\sqrt{t}$-fits. The data density is reduced by a factor of 2500.}
\label{LC_massincrease}
\end{figure}

\begin{figure}[!t]
\centering
\includegraphics*[width=.5 \linewidth]{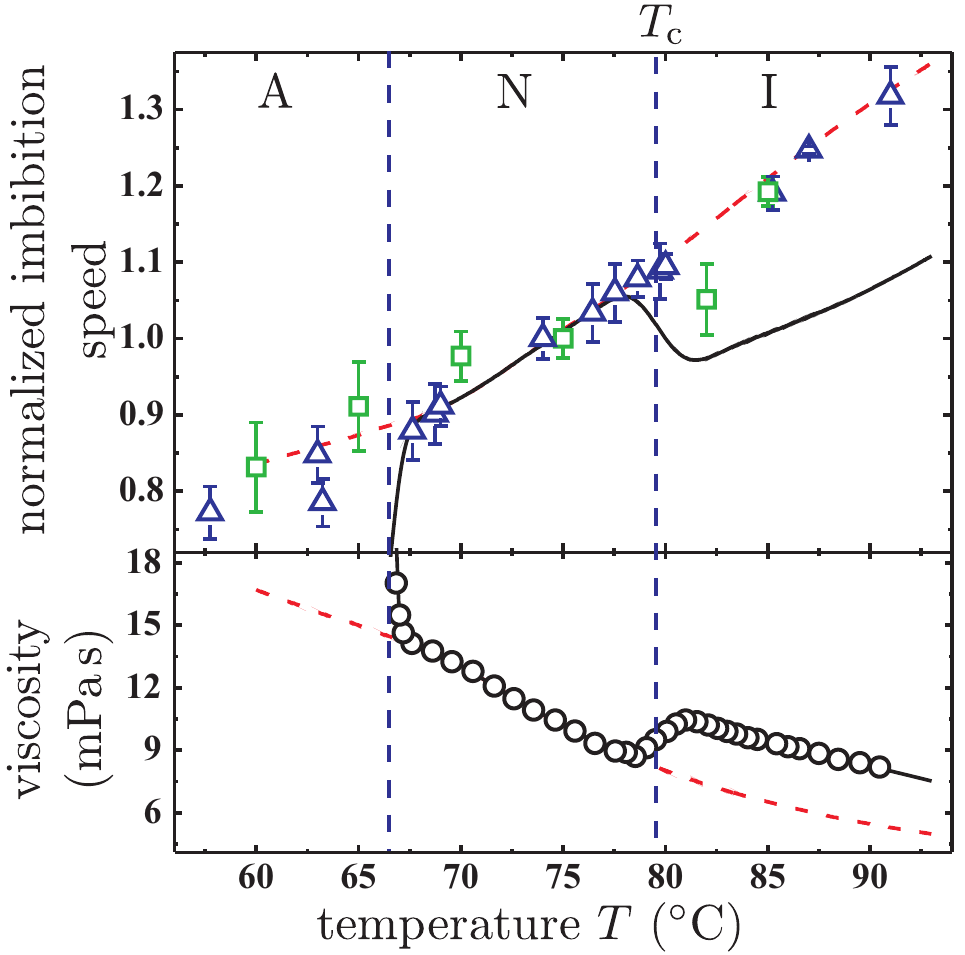}
\caption[Measured normalized imbibition speeds of 8OCB in V5 and V10 as a function of the temperature]{(upper panel): Measured normalized imbibition speeds $v_{\rm n}$ (for $T_{\rm n}=75$\dC) of 8OCB in V5 (${\color{blue} \triangle}$) and V10 ({\color{green} $\square$}), respectively, in comparison with values calculated on the basis of the viscosity values in the lower panel (\Flatsteel). (lower panel): $T$-dependent viscosity of 8OCB by way of comparison (see \bild{LC_visc_8OCB} for a detailed view). The dashed lines correspond to extrapolations of the calculated imbibition speeds and of the viscosity in the absence of the shear viscosity minimum and the presmectic divergence, respectively.}
\label{LC_speeds}
\end{figure} 

The comparison between measurement and prediction reveals a variety of astonishing features. First of all, both V5 and V10 reveal comparable characteristics of the invasion dynamics of the liquid crystal 8OCB. Anyhow, only in the nematic phase the $T$-dependent behavior of the measured imbibition speeds coincides with the prediction based on the $\subs{\eta}{iso}$ values presented in \bild{LC_visc_8OCB}. In particular, the distinctive bump in the proximity of the clearing point $\subs{T}{c}$ as the direct manifestation of the shear viscosity minimum in the theoretical behavior of $v_{\rm n}$ is unambiguously absent. The imbibition speed rather increases monotonously with increasing temperature. 

As mentioned before this distinctive feature of the viscosity at the clearing point is caused by the inset of an alignment of the molecules (in the nematic phase) with respect to the flow direction, that is $\vec{n} \,||\,\vec{v}$. Therefore its absence can intuitively be interpreted in terms of an already existent alignment of the molecules beyond $\subs{T}{c}$. It is obvious to conclude that such an alignment is easily induced by the extreme spatial confinement to cylindrical pores with diameters that are not more than five times the length of the molecule itself. In addition, considering their rigidity and, in particular, the parabolic flow profile established in the pore, it is hard to think of any alternative to the tendency of molecular alignment parallel to the pore axis and, consequently, to the flow velocity $\vec{v}$. From this point of view the absence of the shear viscosity minimum and, consistently, of the nematic to isotropic phase transition is not surprising at all but rather consequent. In this context it is more suitable to label the phase beyond the clearing point not isotropic but {\it paranematic} (P). This term preeminently discloses the solely confinement-induced alignment of the liquid crystal.

It was demonstrated experimentally \cite{Iannacchione93, Dadmun93, Iannacchione94, Crandall96, Cloutier06}, in agreement with expectations from theory \cite{Sheng76, Steuer04, Cheung06}, that there is no `true' I-N transition for liquid crystals confined in geometries spatially restricted in at least one direction to a few nanometers. The anchoring at the confining walls, quantified by a surface field, imposes a partially orientational, that is, a partially nematic alignment of the confined liquid crystal, even at temperatures $T$ far above the clearing point $\subs{T}{c}$. The symmetry breaking does not occur spontaneously, as characteristic of a genuine phase transition, but is enforced over relevant distances by the interaction with the walls.

The assumption of such a paranemtic phase is corroborated by recent birefringence measurements of rod-like liquid crystals (7CB and 8OCB) confined to an array of parallel, nontortuous channels of 10\,nm mean diameter and 300\,\textmu m length in a monolithic silica membrane \cite{Kytik08, Kityk10}. These measurements elucidate that the surface anchoring fields render the bulk discontinuous I-N transition to a continuous P-N transition. The transition temperature $\subs{T}{c}$ is found to be changed only marginally, due to a balance of its molecular alignment induced upward and its quenched disorder (attributable to wall irregularities) induced downward shift. This agrees with the observations of liquid crystals imbibed in tortuous pore networks \cite{Iannacchione93, Dadmun93}.

Interestingly, due to the complete absence of the shear viscosity minimum in the results of the $T$-dependent series of imbibition measurements shown in \bild{LC_speeds} a definition of a P-N transition temperature in confinement is not possible at all. This is confirmed by a simple extrapolation of the $\eta_2$ viscosity to higher $T$, which saliently reproduces the measured imbibition speeds beyond the bulk clearing point. This can only be interpreted in terms of an extremely high degree of orientation already existent in the paranematic phase; at least higher than suggested by the birefringence measurements \cite{Kytik08}. This difference is most probably due to the basically differing detection method of the P-N transition: normally birefringence or calorimetry (DSC) measurements performed with the confined {\it static} liquid are applied for this purpose. But, the viscosity measurements presented here refer to the liquid's dynamics in mesopore confinement. As already mentioned before, the additional emerging flow velocity and in particular the velocity gradient seemingly enhance the paranematic orientational alignment in the $\eta_2$-configuration significantly.

The second remarkable feature of the results presented in \bild{LC_speeds} is the absence of the presmectic divergence of the viscosity of the freely flowing liquid crystal. This would result in a dramatic drop of the imbibition speed due to the inset of smectic layering. However, the measured values do not indicate such an effect. Its absence rather suggests a suppression of the A phase in favor of the N phase. This is elucidated by a simple extrapolation of the viscosity to lower $T$ as indicated in \bild{LC_speeds}, which preeminently reproduces the measured imbibition speeds.

What are the reasons for this discovery? First of all, the confinement to a cylindrical pore (rather than to a film geometry like in the Couette flow depicted in \bild{LC_visc}) renders the $\eta_3$ viscosity as unfavorable as the $\eta_1$ viscosity. This again clarifies the high stability of the nematic $\eta_2$-configuration in confinement. A cross-over behavior as ascertained for the freely flowing bulk liquid can hence be excluded. Yet, the presmectic divergence of $\eta_2$ is caused by the destruction of precursors of smectic planes for $\vec{n} \,||\,\vec{v}$. From this point of view the suppression of the A phase is not surprising at all but a mere consequence of the overall stabilization of the $\eta_2$-configuration in the mesopore confinement. 

Even the static mesopore-confined liquid crystal shows such modified mesophase behavior \cite{Iannacchione94}. For example the heat capacity anomaly typical of the second-order N-A transition in rod-like liquid crystals immersed in aerogels is absent or greatly broadened \cite{Bellini01, Qian98}. Nuclear magnetic resonance (NMR) measurements revealed the lack of pretransitional smectic layering due to the rough surface of the confining walls \cite{Crawford93}. Furthermore, a systematic study of the influence of the degree of confinement indicates that the N-A transition becomes progressively suppressed with decreasing pore radius whereas the stability range of the nematic phase is increased \cite{Kutnjak03, Kityk10}.

Finally we'd like to comment on results from an examination on the absolut invasion dynamics of the liquid crystal as opposed to the analysis of normalized imbibition coefficients just presented. This investigation was performed according to the procedure introduced in section~\ref{confinedwater} referring to mesopore-confined water. For the liquid crystal 8OCB we determined slip lengths of $b=(-1.11\pm 0.23)$\,nm for V5 and $b=(-1.54\pm 0.31)$\,nm for V10, respectively. Referring to the molecular shape stated earlier we may hence, again, conclude that one layer of flat lying molecules is pinned to the pore walls and does therefore not contribute to the flow dynamics. Apparent velocity slippage at the walls, as was expected to be associated with the molecular alignments in the channels \cite{Heidenreich07}, could not be detected in our study.

\section{Conclusions}
After a short introduction of liquid imbibition in porous hosts we have presented a gravimetric study on spontaneous imbibition of water and the liquid crystal 8OCB. We inferred from our studies an interfacial boundary layer adjacent to the pore walls with a defined thickness whose dynamics are mainly determined by the interaction between liquid and substrate. This manifests itself in terms of a negative velocity slip length for both liquids investigated. The flow properties of the pore-condensed molecules in the pore center are, however, remarkable robust, that is bulk-like or at least similar to the bulk phase (in the case of the liquid crystal).  

The $T$-dependent imbibition measurements of the liquid crystal 8OCB revealed that confinement plays a similar role as an external magnetic field for a spin system: the strong first-order I-N transition is replaced by a weak first-order or continuous paranematic to nematic transition, depending on the strength of the surface orientational field \cite{Stark02}. Based on detailed knowledge of the static (equilibrated) liquid's behavior in the mesopores as deduced from previously conducted birefringence experiments, we were able to procure complementary results with respect to its dynamic (non-equilibrium) behavior. The additional emerging flow velocity and in particular the velocity gradient enhance the paranematic orientational alignment significantly rendering the P-N transition even broader than known from the equilibrium state. Due to the high stabilization of the $\eta_2$-configuration in the N phase the A phase is suppressed and the stability range of the nematic phase is increased. 

The finding of two distinct species (with regard to flow dynamics) is reminiscent of the partitioning of pore condensates found in vapour sorption isotherms (film-condensed versus capillary-condensed state) \cite{Huber99}, of the two species often reported in studies on self-diffusions dynamics (slowed-down or increased self-diffusion versus bulk dynamics) \cite{Baumert2002, Koppensteiner08, Kusmin10a, Kusmin10b}, of the effects of pore condensates on the deformation of the matrix (expansion versus contraction of the pore walls upon change from film-condensed to capillary-condensed state) \cite{Guenther2008}. Last but not least, it reminds of the distinct crystallization behaviour of pore condensates (an amorphous boundary layer versus a crystallized fraction of molecules in the pore center) \cite{Huber99}. In principle, this partitioning, but also the highly stabilized nematic phase of a confined liquid crystal, can be traced to the strong influence of the substrate potential on the liquid or solid layers, respectively, right adjacent to the pore walls \cite{Scheidler00, Scheidler02, Klapp02}. 

From a more general perspective, our experimental studies demonstrate that an understanding of the non-equilibrium behavior of confined soft-matter system necessitates a thorough understanding of their equilibrium behavior, which quite often differs markedly from the macroscopic phenomenology \cite{Binder08, Shen10}.

\section*{Acknowledgments}
It is a great pleasure to thank Andriy Kityk, Klaus Knorr, Dirk Wallacher, and Howard A. Stone for stimulating discussions. Financial support by the DFG under grants Hu850/2 (1-2) within the priority program `Nano- and Microfluidics' is acknowledged.

\end{document}